\documentclass{article}
\PassOptionsToPackage{numbers, sort, compress}{natbib}

\usepackage[preprint]{neurips_2021}

\usepackage[utf8]{inputenc} 
\usepackage[T1]{fontenc}    
\usepackage{hyperref}       
\usepackage{url}            
\usepackage{booktabs}       
\usepackage{amsfonts}       
\usepackage{nicefrac}       
\usepackage{microtype}      
\usepackage{xcolor}         

\usepackage[pdftex]{graphicx}
\usepackage{amsmath}

\usepackage{siunitx}
\usepackage{graphicx}
\usepackage{lastpage}
\usepackage{enumerate}
\usepackage{fancyhdr}
\usepackage{mathrsfs}
\usepackage{xcolor}
\usepackage{listings}
\usepackage{float}
\usepackage{comment}
\usepackage{framed}
\usepackage{tikz}
\usepackage{caption}
\usepackage{subcaption}
\usepackage{enumitem}
\setitemize{noitemsep,topsep=6pt,parsep=4pt,partopsep=0pt}

\usepackage[ruled,vlined,linesnumbered]{algorithm2e}
\SetKwInput{KwInput}{Input}

\title{Optimizing piano practice with a utility-based scaffold}
\pdfoutput=1    
\author{
  Alexandra Moringen\thanks{Correspondence to  \texttt{abarch@techfak.uni-bielefeld.de}} \\ 
  University Bielefeld \\
   \and
   S\"oren R\"uttgers  \\
   University Bielefeld  \\
   \and
   Luisa Zintgraf  \\
   University of Oxford \\
   \and
   Jason Friedman \\
   Tel Aviv University \\
   \and
   Helge Ritter \\
   University Bielefeld \\
}

\begin{document}

\maketitle

\begin{abstract}
  A typical part of learning to play the piano is the progression
  through a series of practice units that focus on individual
  dimensions of the skill, such as hand coordination, correct posture,
  or correct timing.  Ideally, a focus on a particular practice method
  should be made in a way to maximize the learner's progress in
  learning to play the piano.  Because we each learn differently, and
  because there are many choices for possible piano practice tasks and
  methods, the set of practice tasks should be dynamically adapted to
  the human learner. However, having a human teacher guide individual
  practice is not always feasible since it is time consuming,
  expensive, and not always available. Instead, we suggest to optimize
  in the space of practice methods, the so-called practice modes.  The
  proposed optimization process takes into account the skills of the
  individual learner and their history of learning.  In this work we
  present a modeling framework to guide the human learner through the
  learning process by choosing practice modes that have the highest
  expected utility (i.e., improvement in piano playing skill).  To
  this end, we propose a human learner utility model based on a
  Gaussian process, and exemplify the model training and its
  application for practice scaffolding on an example of simulated
  human learners.
\end{abstract}

\section{Introduction}
Learning to play sports or to play a musical instrument is not a
trivial task, and normally we cannot learn these skills by simply
observing an expert. Rather, they are learned with a help of a teacher
who provides scaffolding. ``Scaffolding is a reciprocal feedback
process in which a more expert other (e.g., teacher, or peer with
greater expertise) interacts with a less knowledgeable learner, with
the goal of providing the kind of conceptual support that enables the
learner, over time, to be able to work with the task, content, or idea
independently''~\cite{Renninger2012}. The main contribution of this
work is a model of a scaffold that guides the human learner through
the process of learning to play the piano by choosing the practice
method that has the highest expected utility (i.e., improvement in
piano playing skill).  To this end, we propose a human learner utility
model based on a Gaussian process, and exemplify the model training
and its application for practice scaffolding on an example of a
simulated human learner.

The mastery of the learner increases as they become more proficient
with a growing mixture of abilities.  Playing the piano is a typical
complex combinatorial skill which builds upon gradual mastery of
prerequisite skills \cite{griffith18}. Skilful performance requires a number of
abilities, including being able to read notes, follow the speed,
produce correct rhythmic structures, use a correct hand posture,
and coordinate movements of both the hands and feet. In a band, an
additional ability to coordinate with others adds to the challenge.  A
typical example of the scaffolding process is reducing the complexity
of the task by reducing the degrees of freedom~\cite{Zydney2012ESL},
and sometimes even by off-loading the rest onto something or somebody
else~\cite{Pea2004STD}.  In piano practice, it is common to first
practice with each hand individually.  Here, the part of the
non-practicing hand can be offloaded onto the computer.  In the above
example, a good level of mastery for both hands separately is
necessary in order to be able to concentrate on the task of
coordinating both hands.  The focus on practicing one or several
aspects or modes of the complex skill by reducing the complexity of
the overall task accordingly is termed here \textbf{practice
  mode}. For the generation of practice modes, we need a set of
procedures to simplify the target task for practice, to decompose it
to a subset of modes. An example of this is a practice mode that
targets improving timing in a particular piece of score\footnote{by
  ``score'' we mean the music score.}, by substituting every note to
one single pitch, such as middle C. This reduces the complexity of the
task and enables the learner to just concentrate on timing
correctness, instead of both timing and pitch correctness.  Another
practice mode that targets improvement in pitch correctness allows the
learner to play as slowly as they want, waiting for the learner to
play the next note for as long as they need before moving to the next
note. This enables the learner to concentrate on pitch correctness
only and not think about playing with correct timing. In general, a
practice mode may need to specify both the target task adjustments (as
in timing practice mode) as well as the method of practice
organisation (such as waiting for the learner to play the correct note
in pitch practice mode).

How to guide or even accelerate learning of multimodal skills such as
playing the piano is an open research question. Although a large
amount of research has been dedicated to the educational domain and
towards creating intelligent human-in-the-loop tutoring systems
\cite{koedinger2013new,lee2012impact,rafferty2011faster}, very little
work exists for optimizing practice of motor tasks, such as piano
playing. While research efforts focus mainly on analysis and
rehabilitation of professional musicians~\cite{Kita2021ACC}, or the 
differences between the professional player and the novices
~\cite{Furuya2007RPS}, no recent work targeted scaffolding of learning
for non-professional piano players.

Our long-term goal is to design a framework for intelligent tutoring
with the human in the loop, to help the human learner (HL) stay
engaged, get suitable challenges, and receive constructive feedback on
their performance.  In this work we propose a scaffolding framework
consisting of a teacher-complementing model infrastructure and an
experimental setting, both controlling the learning process. In particular, we train a Gaussian Process
(GP)~\cite{Rasmussen2006GPM} to represent the utility of practice
modes, and illustrate the results in a simulated experiment.  The
final application performs an optimized selection of practice modes
for a HL on one complexity level.

Due to the fact that our application should be suitable to scaffold a
real human learner, a case in which data acquisition is costly, we
pursue a modeling approach with a GP that enables us to model data with very few
samples.  GPs
provide not only the approximation in the functional space but also
 can represent the uncertainty about the estimate at a particular
point, which is a useful property for both data-acquisition and
scaffolding.
Following previous work~\cite{Mu2020TSA}, we build on the assumption
of a static tree-like curriculum (described in more detail in
Section~\ref{sec:curriculum}). In~\cite{Mu2020TSA} the authors propose
to use the so-called actionable features to dynamically react to the
learner's needs.  An actionable feature encodes an intervention,
e.g., that the learner needs more practice on a corresponding
prerequisite skill before moving up the curriculum tree.  In this work
we extend the concept of actionable features as described in ~\cite{Mu2020TSA} to practice modes (see
Section~\ref{sec:notation} for a more detailed description).  This new concept can be illustrated  particularly well 
on an example of  piano learning domain. Here we can track
far more information characterizing performance of the learner and the amount of 
improvement, in comparison to typical educational applications
that in some cases only output whether the question has been answered
correctly or not.  Due to a clear multimodal structure of the piano
playing task, the practice modes can be derived from the task
automatically based on expert advice for a wide range of pieces
invariant to the complexity level. This is different to the 
educational applications, where new actionable features might be
necessarily on new complexity levels. In this work we introduce the
basic modeling infrastructure needed to optimize practice mode
selection for a \textit{piano playing scaffold}, and the simulated
learner environment that provides us with insights about how to shape
future human learner studies.

\section{Components and Notation}
\label{sec:notation}
\begin{figure}[h!]
  \centering
  \includegraphics[height=5cm]{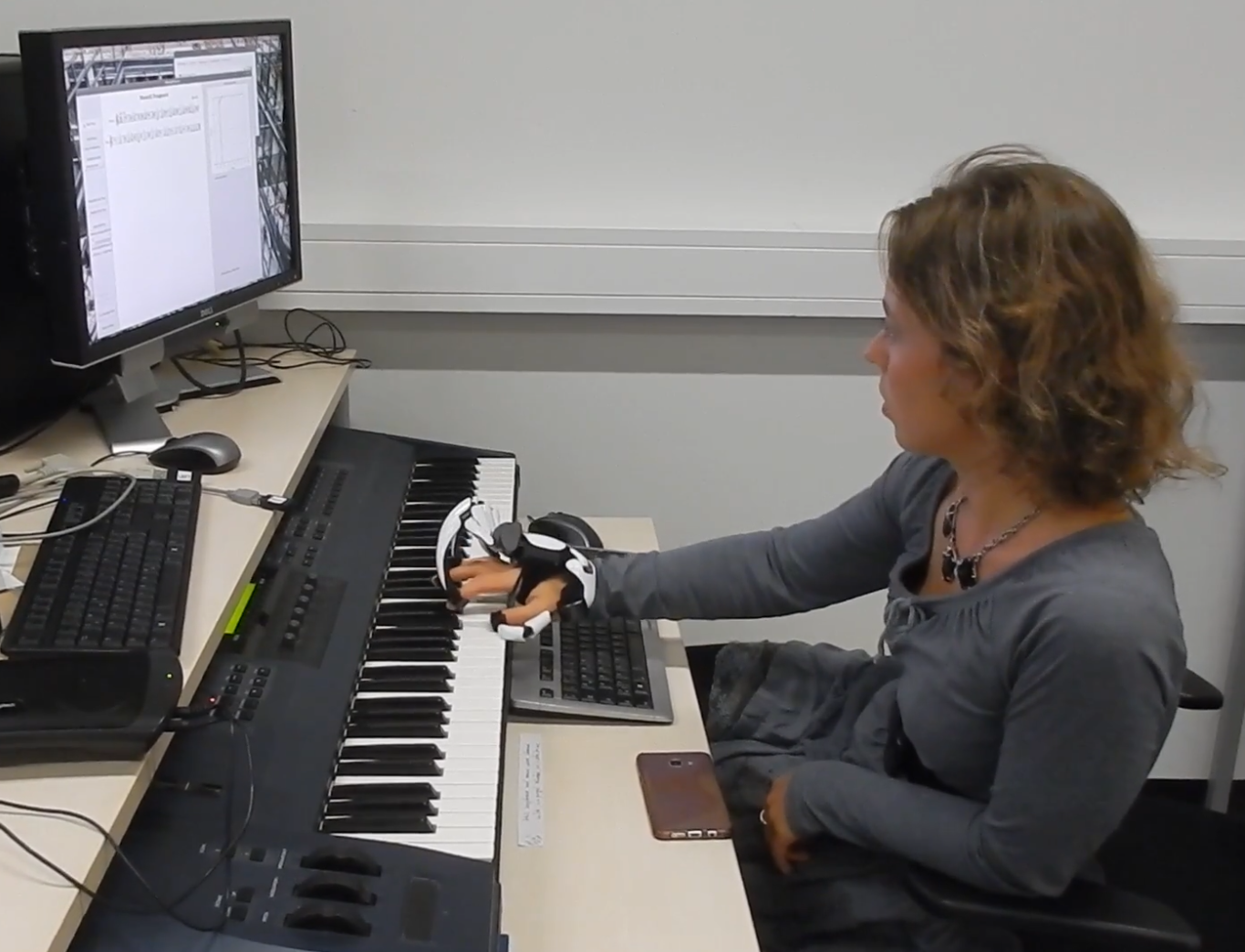}
  \includegraphics[height=5cm]{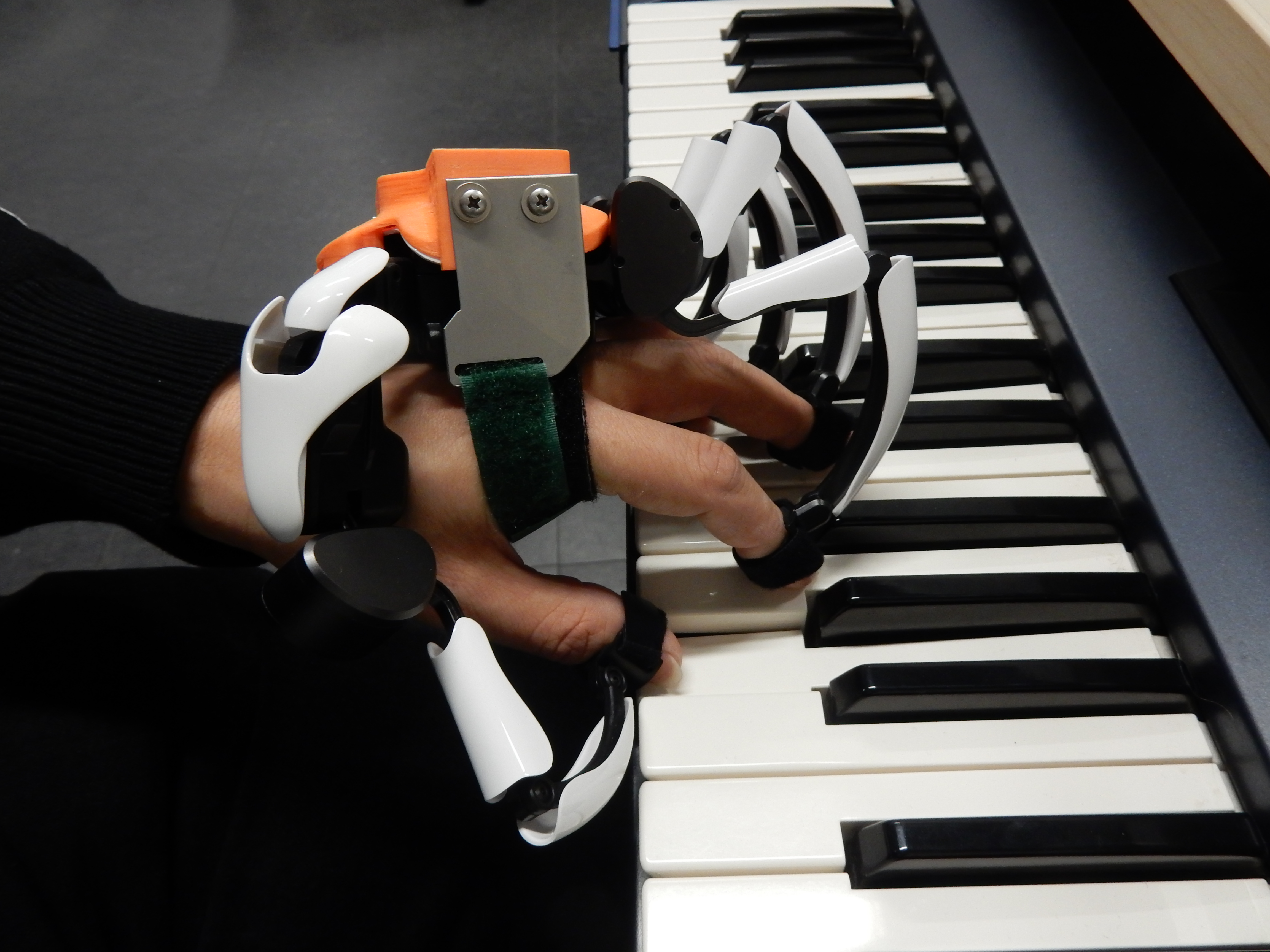}
  \caption{\label{fig:dexmo} (Left:) The envisioned intelligent
    tutoring setup consisting of a GUI displaying a music score, a keyboard 
    and a practice mode specified by guidance with an exoskeleton Dexmo. (Right:)
    an example of a practice mode implemented through guidance with
    Dexmo. During this practice mode, Dexmo generates a
    sequence of force impulses corresponding to the structure of the
    score, and moves the corresponding fingers of the human learner
    (HL) according to the start and end of the note. }
\end{figure}

\begin{itemize}
  
\item \textbf{Tasks} $\tau$.  Mathematically, a single task is denoted
  $\tau\in\mathbb{R}^N$. A task is defined as a vector that represents
  a set of $N$ task attributes (playback speed in beats per minute,
  number of notes, rhythmic complexity, volume, number of hands, etc.)
  Task attributes may also specify the performance method or goal,
  such as focus on correct timing and pitch, etc.  We differentiate
  between a target task $\tau^{\text{target}}$ (\textit{coll.}
  ``parent task'') and a set of corresponding \textbf{practice modes}
  $\tau_M$ that are derived from the target task by using a subset
  $M\subseteq\{1,\ldots,N\}$ of target task attributes.  A practice
  mode is then represented by an $N-$dimensional vector with irrelevant
  task attributes zeroed out; we  normalize  the relevant
  task attributes to values $>0$. Importantly, all
  tasks and all practice modes have the same dimensionality.

  Example: let $\tau^{\text{target}}$ be a task, which is
  characterised by a full set of task attributes. The performance of
  the target task is tracked in both hands w.r.t. pitch, timing, and
  posture correctness.  We generate a practice mode by selecting
  attribute values, e.g. $(\tau_i, \ldots, \tau_k)$ that correspond to
  practising aspects of the target task such as pitch correctness separately
  for each hand, clapping instead of playing to improve timing, or being guided by an
  exoskeleton to improve fingering (see Figure~\ref{fig:dexmo}).

\item \textbf{Task set} $\mathcal{T}$. We denote a (finite) set of
  tasks as $\mathcal{T}$. We specify a partition of the task set $\mathcal{T}
  =\bigcup \mathcal{T}_c$ into subsets of different complexity
  levels, denoted by $c$ (using prior knowledge
  from,  e.g., a textbook or app).

\item \textbf{Task distributions} $p$. We denote a task distribution
  as $p_\phi^\mathcal{T}$, with support over the tasks in the set
  $\mathcal{T}$ and parameterised by $\phi$. (E.g., $p$ could be
  Gaussian and $\phi$ the mean and variance of that Gaussian). We can
  sample individual tasks from this distribution,
  $\tau\sim p^\mathcal{T}_\phi$. Each complexity level
  $c$ is characterised by its own distribution over the task-subset, denoted $p_\phi^{\mathcal{T}_c}$.
\item \textbf{Music score generator $G$}. $G$ is a mapping of task
  attributes to a corresponding score $G:\tau\rightarrow s$, where $s$
  is a music score (which is a sheet of notes the student has to
  play). The music score generator generates scores for both full
  tasks, or the practice modes according to the attributes specified by $\tau$.

  For example, it could use
  attributes such as number of active hands (left, right, both),
  number of active fingers (1-5), rhythmic complexity (usage of
  half, quarter notes, etc.) or the note range (e.g. C through G,
  just the natural (white) keys, an octave, the whole scale) to either
  select a piece from a database or to construct it randomly to
  satisfy the above-mentioned constraints. An example of a randomly
  generated task for one hand in the range of a whole octave is
  illustrated in Figure~\ref{fig:gui} (Appendix
  \ref{appendix:piano_gui}).  Randomly generating practice tasks
  allows us to produce a wide range of practice opportunities that can
  be individually tuned to the skill level of the HL and whose
  complexity may increase in a fine-grained manner. The underlying
  data structure is visualized in Appendix~\ref{appendix:task_generation}.
 
  \item \textbf{Performance error} $\rho(\tau) \in\mathbb{R}^K$. 
  The performance error of a human learner on a task $\tau$ evaluated
  w.r.t. a $K$-element set of attributes  and denoted by 
  $\rho(\tau)$.  Performance error can be evaluated 
  w.r.t.  pitch correctness, timing correctness,
  correctness of hand posture, pressure, fingering, volume, etc.
  We here assume
    for simplification that we can evaluate the performance error level for
    each attribute with a single scalar value (if this is a
    restriction, we can always split an attribute into several
    "subattributes", one for each required performance error dimension).

  \item \textbf{Human learner}: We assume that each HL $h_a: \tau
    \rightarrow \rho_\tau$ has a set of attributes $a \in
    \mathbb{R}^T$ characterizing their personal talents, strengths and
    weaknesses.  In Section~\ref{sec:simulation} we differentiate
    between three simplified HL groups defined by  their
    performance error dynamics w.r.t. timing and pitch.

 \item \textbf{Task utility function} $g$: For a given task $\tau$ we have a set
   of practice modes  $\tau_m \in \tau_M$, where
   $M\subseteq \{1,\dotsc,N\}$ is a subset of attribute indices of the
   task.  The Gaussian process represents the following utility
   function of selecting a practice mode for the human learner:
  \begin{equation}
       \label{eqn:impact}
       g_a: \tau_m  \rightarrow u_{\tau}, 
      \end{equation}
where the utility $u_{\tau}$ of the practice mode  $\tau_m$ for reducing performance error of $\tau$ 
is approximated by the error delta after one practice with $\tau_m$:
\begin{equation}
u_{\tau} = \rho^{\text{pre}}(\tau_m) - \rho^{\text{post}}(\tau_m).
\end{equation}

Such approximation  is necessary due to the fact that it might be impossible for a HL to
initially play the target task without performing multiple practice
modes first. In this case, we cannot initially let HLs perform the target task
directly.
 \item \textbf{Scaffold} $S$: takes predictions of the task utility
   function as an input and selects the practice mode with the highest
   utility.  Building upon the definition of scaffolding, in future
   work we will extend $S$ to increase or decrease the task complexity
   according to the progress of the HL.
\end{itemize}    

\begin{figure}
     \centering
     \begin{subfigure}[b]{0.45\textwidth}
         \centering
         \includegraphics[height=6.5cm]{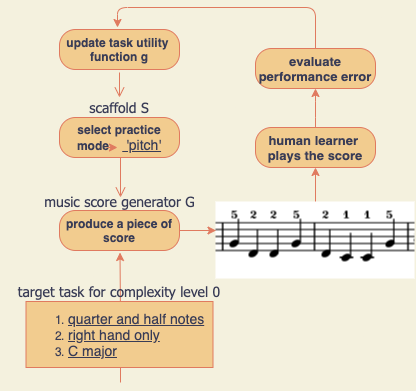}
         \caption{Interaction between the scaffold and the learner.
           For a target task and a practice mode specification, the
           music score generator $G$ produces a piece of score. In
           this case the practice mode with the highest utility
           targets improvement of pitch. The error of the learner is
           evaluated and is employed for the update of the utility
           function. }
         \label{fig:interaction}
     \end{subfigure}
     \hfill
     \begin{subfigure}[b]{0.45\textwidth}
         \centering
         \includegraphics[height=6.5cm]{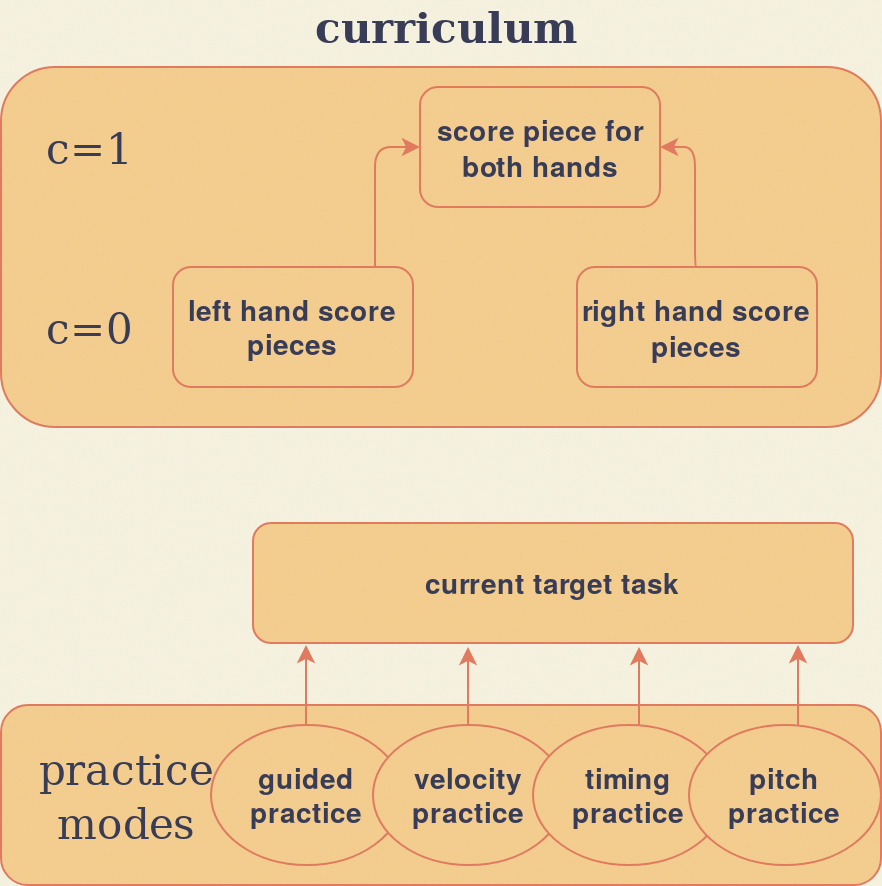}
         \caption{(Top:) Curriculum tree example.  The tasks on each
           complexity level are generated by the music score generator
           $G$ according to the specifications provided by the
           curriculum within each tree node. (Bottom:) a set of
           potential practice modes for the current target task. The
           most useful practice modes are selected by the scaffold
           $S$ for a HL $h_a$. }
         \label{fig:curriculum}
     \end{subfigure}
     \caption{Practice infrastructure for the human learner. }
\end{figure}

Figure~\ref{fig:interaction}  exemplifies main stages of one loop iteration describing the reciprocal interaction between the human learner and the scaffold.

\section{Practice Infrastructure for the Human Learner}

In this section, we first outline the individual parts of the 
practice infrastructure for the human learner, and then focus our
attention on the parts that can be optimized. 

\subsection{Practice Infrastructure Components}
\label{sec:curriculum}
Building upon previous work on intelligent tutoring systems (e.g.~\cite{Mu2020TSA}),  as well as the theory of scaffolding coming from the educational psychology~\cite{Renninger2012}, we
assume that the practice infrastructure consists of two main parts: 1)
a static curriculum represented by a tree-like structure of
prerequisite skills, and 2) a scaffold whose role is, based on the
utility function $g$ and a target task $\tau^{\text{target}}$, to
dynamically select practice modes for the HL depending on their
estimated utility.

On the one hand, a static curriculum is a learning plan anchored in
music-theory, with the human learner being gradually introduced to
tasks with increasing complexity. Examples are tonal keys of
increasing number of sharps (C major, G major, D major), moving from
playing single notes, then intervals, then chords, moving from playing
adjacent notes to extending the spatial distance between the keys,
moving from simple time signatures such as 4/4, to 3/4 to more
difficult ones such as 5/4, etc.  On the other hand, the scaffold is the
interactive component that takes the learner's skill level and their
individual practice utility into account to pick the most useful
practice modes on-the-fly.  Figure~\ref{fig:curriculum} 
illustrates both components: the interactive (bottom) and the static
(top), respectively.  The top row shows an example for a
curriculum consisting of two complexity levels. On the complexity
level $c=0$ there are tasks to practice for one hand only. On the
level $c=1$ the generated task requires both hands to play
simultaneously. One can only move up to $c=1$ after mastering both
prerequisite skills for $c=0$. It is intuitively clear that it is
impossible for a novice to play a piece in which they have to
coordinate both hands playing different parts before they have
reached a sufficient degree of proficiency in playing each hand
individually. For example, the learner needs to be able to hit approximately
the right notes while keeping track of the rhythmic structure in
the left and right hands, individually, before doing the same for
both hands.  

For a given target task, the goal of the scaffold is to select the
most useful practice mode such as one targeting correct pitch or
correct timing  to enable the HL to master it as quickly
as possible.  The bottom part of Figure~\ref{fig:curriculum} shows
several possible practice modes targeting correct pitch, correct
timing, increasing the playback speed, or guidance by a exoskeleton.

\subsection{Learning the Task Utility Function }
A key element in the present scheme is the utility predictor function $g(\cdot)$, which
is the adaptive part in our model. It is represented by a Gaussian process
 (GP; see
Section~\ref{sec:simulation}), which is specified by a mean function
$m(\cdot)$ and a kernel $k(\cdot,\cdot)$, and is denoted as 
  $g(\cdot) \sim \mathcal{GP}(m(\cdot), k(\cdot,\cdot))$. 
In this work we explore a GP for optimal selection of
practice modes for randomly generated tasks on one complexity level.
As a first step, we show the tests
conducted in simulation that trains a GP model to generate a practice
mode policy that results in the highest utility, currently defined  by the reduction of the  performance error caused by practice with the respective mode.
We chose a zero-mean function and a Matern kernel ($\nu = \frac{5}{2}$). The
kernel features two parameters \textit{lengthscale} $l$ and
\textit{variance} $\sigma^2$ and is denoted as follows: $C_{5/2}(d) = \sigma^2\left(1+\frac{d}{l}+\frac{d^2}{3l^2}\right)\exp\left(-\frac{d}{l}\right)$. The parameters will be the
same for all input dimensions, as is typical for most GP
implementations. Therefore,  task attributes from different value ranges  have
to be normalized or otherwise adapted prior to the GP training. For those
task attributes that simply can not be projected onto a linear scale
(e.g. the use of left, right, or both hands) a one-hot encoding needs
to be used. This does not mean that there has to be no covariance
between each category, as the one-hot encoding can be scaled
($[0,0.2]$ instead of $[0,1]$), which results in each category having
the same covariance as each other category. Other attributes where
some categories are more similar to one another than to others can be
encoded using handcrafted multi-dimensional embeddings. 
The practice modes themselves are an example of such categorical input. In
this specific case we do not want any covariance between each
category, which could be realized by using a high scaled one-hot
encoding, or since the covariance between points is zero regardless,
embed them in a single dimension by placing them sufficiently far
apart.

\section{Simulated Experimental Setting }
\label{sec:simulation}

\begin{algorithm}[H]
	\label{alg:ts_error_valc}
	\SetAlgoLined
	\KwInput{HL-model (\textit{performer}), noise\_var, bpm, note\_range, practice\_mode}
	\KwResult{utility}
	initial\_error.timing = $\epsilon$ * $\frac{\text{bpm}}{10}$   \hfill $\epsilon \sim \mathcal{N}(1,\text{noise\_var}^2)$\\
	initial\_error.pitch = $\epsilon$  * $\begin{cases}
      0.5 & \text{if } \text{note\_range} = 0 \\
      1.5 & \text{if } \text{note\_range} = 1 \\
      3   & \text{if } \text{note\_range} = 2
    \end{cases}$ \hfill $\epsilon \sim \mathcal{N}(1,\text{noise\_var}^2)$\\

    error\_pre = initial\_error.copy()\\
    \If{performer == bad\_pitch}{
    	error\_pre.pitch = initial\_error.pitch $\times$ 1.75
    	}
    \If{performer == bad\_timing}{
    	error\_pre.timing = initial\_error.timing $\times$  1.5
    	}
    
	\texttt{// calculate utility for a given practice mode } \\ 

    \If{practice\_mode == improve\_pitch}{
    	error\_post.pitch = error\_pre.pitch $\times$ 0.5
    	}
    \If{practice\_mode == improve\_timing}{
    	error\_post.timing = error\_pre.timing $\times$ 0.5
    	}    
    	
    utility = error\_pre.timing - error\_post.timing + error\_pre.pitch - error\_post.pitch\\
    utility *= $\epsilon$ \hfill $\epsilon \sim \mathcal{N}(1,\text{noise\_var}^2)$\\
    utility -= mean\_utility \hfill \textit{(mean\_utility is a predefined constant)}\\

    \Return{utility}
	 \caption{Pseudocode of the error and utility calculation}
\end{algorithm}

In order to test the practice infrastructure without 
performing trials with HLs, similar to \cite{Mu2020TSA} we build a
simulated experimental setting consisting of the following building
blocks:
\begin{enumerate}
\item The Gaussian process representing the utility function: for a set of
  task parameters, including  a specification of the
  practice mode, it approximates the corresponding utility. The mode
  with the highest expected utility then gets selected.
\item Two simulated practice modes denoted by \texttt{IMP\_TIMING} and
  \texttt{IMP\_PITCH} that model optimal practice that improves timing and
  pitch correctness, respectively.
\item A two-dimensional task specification consisting of target
  playback speed denoted by beats per minute (bpm) and three
  categories for the range of notes that are employed for task
  generation, with $\text{bpm} \in \{50, \ldots, 200\}$ and
  $\text{note-range} \in \{ 0, 1 ,2 \}$. Note range parameter exemplifies  a 
  categorical input attribute, and will in future determine 
   the number and placement of notes employed  for the music score generation.
  All
  generated tasks specified by $(\text{bpm}, \text{note-range})$ tuples  are
  located on the complexity level $c=0$.
\item Three simulated groups of HLs, $h_1-h_3$, characterized by
  different error dynamics (see Algorithm~\ref{alg:ts_error_valc}).
  $h_1$ has difficulties playing the right pitch, $h_2$ has
  difficulties with the timing, while $h_3$ features no such error
  amplification (balanced case).  As a coarse model of the learning
  effect of the practice mode we simply assume that working through a
  practice mode reduces the associated error measure by some fixed
  percentage, e.g. $50\%$. The utility of a practice mode is then defined by the
  decrease of error in the post-practice-mode performance (see
  Algorithm~\ref{alg:ts_error_valc}).
  The scaling factors chosen in lines 5 and 8  of the algorithm are
  designed to generate different optimal ground truth  practice mode policies for each HL group (discussed in more detail below).
\end{enumerate}
The experiments showed that leaving the bpm input-value not normalized
leads to very little improvement of the model with the increasing
number of training iterations. Learning is much faster with normalized
inputs (therefore we use an empirically estimated
$\frac{\text{bpm}}{10}$). We subtract a mean\_utility from the
calculated utility value to better fit the  GP model assumption
(zero mean function $m(\cdot)$).

	\begin{algorithm}[H]
	\label{alg:ts_experiment}
	\SetAlgoLined
	\KwInput{The learner\_performance\_error and calc\_utility functions}
	\KwResult{The policy of the trained GP }
	 c = 0\\
	 GP = GP()\\
	 \For{\_ in num\_iter}{
	   tp = random\_task\_parameters()\\
	   error\_pre = learner\_performance\_error(tp)\\
	   
	   practice\_mode = GP.get\_practice\_mode(c, tp)\\
	   error\_post = learner\_performance\_error(tp, practice\_mode)\\
	   
	   utility = calc\_utility(error\_pre, error\_post)\\
	   GP.add\_data\_point(c, tp, practice\_mode, utility)\\

	 }
	 \caption{Pseudo code of the GP training.}
	\end{algorithm}

Therefore, practice by a HL from a group $h_a$ 
contributes one datapoint for the training
of the Gaussian utility predictor model $g_a$. This datapoint has as its input
elements $(\text{bpm}, \text{note\_range}, \text{practice\_ mode\_type})$. The target output in the datapoint is the actual 
utility value inferred from the training result. 
Our goal is to check how well the Gaussian model can capture this
setting from few datapoints.

\begin{figure}
\centerline{\includegraphics[width=0.75\columnwidth]{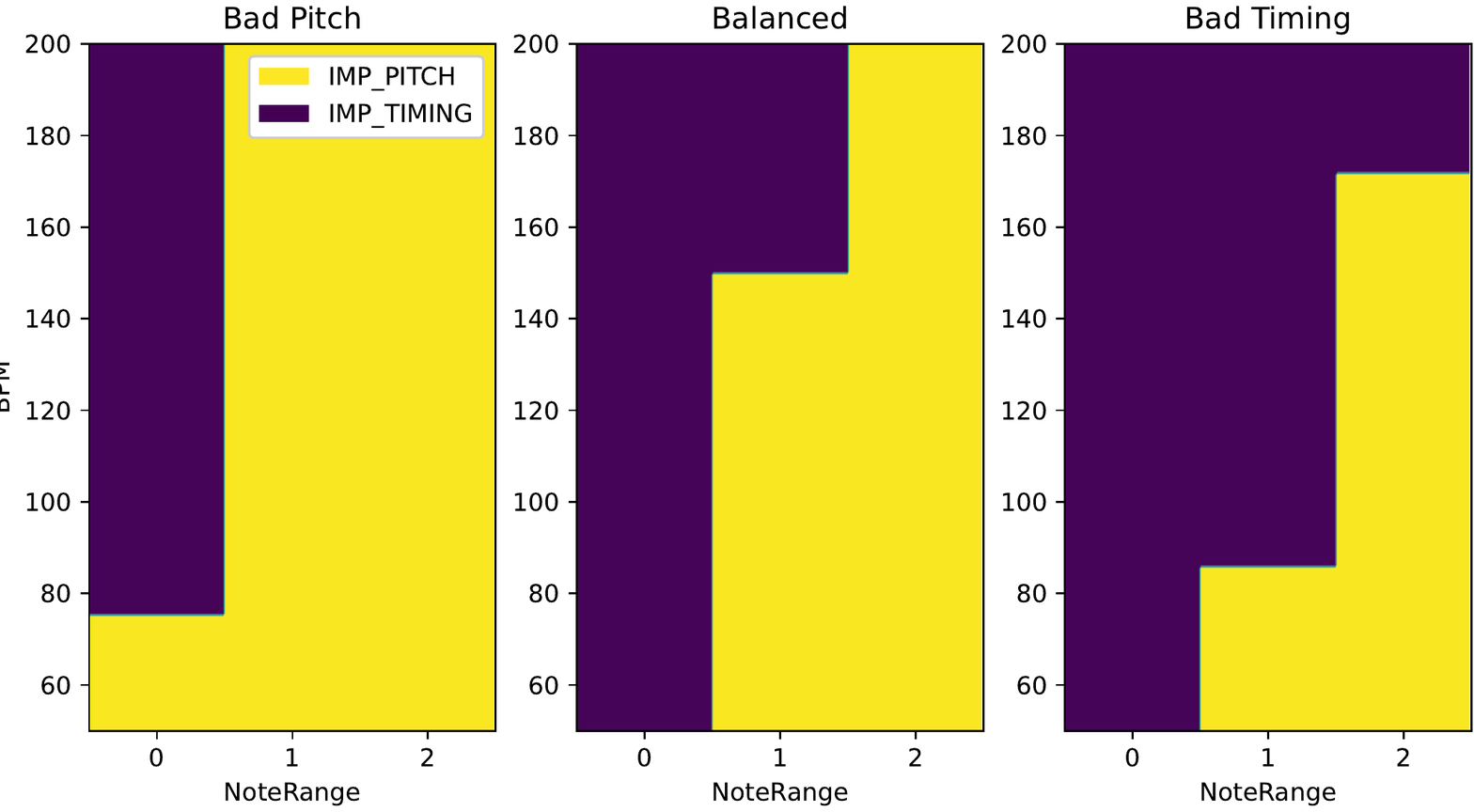}	}
\caption{Optimal \textit{ground truth} practice mode policy for each
  considered task-parameter combination and each HL group. Three
  subfigures correspond to three \textit{HL} group: bad pitch (left), a
  balanced error (middle), bad timing (right). Different types of
  error dynamic determine HL-specific ground truth utility for each
  parameter combination, and, finally, result in a different practice
  modes being optimal in each one of the three HL groups. Color
  yellow depicts an optimal selection of \texttt{IMP\_ PITCH} practice
  mode, and color purple corresponds to \texttt{IMP\_TIMING} practice mode. }
\label{fig:ts_optimal_policies}
\end{figure}

To be able to evaluate the quality of our model, we need to compare
the GP-based policy with an optimal policy.  Using the simple model assumptions
outlined in above points 1.-4. we can calculate the ground truth for
the optimal policy that specifies the best practice mode to suggest
for each combination of task parameters (see
Figure~\ref{fig:ts_optimal_policies}).

The pseudocode of the GP training is shown in Algorithm
\ref{alg:ts_experiment}.  In each iteration, random task parameters
get generated within their respective bounds. We then let our HL-model
``play'' the task and get the pre-practice error. This is a
multi-dimensional error with its timing dimension determined by the
tempo (bpm) and the pitch dimension determined by the value of the
note-range parameter. These errors are increased according to the
selected HL-model $h_1$, $h_2$ or $h_3$, respectively. We then use the
utility predicted by the GP to select the best practice mode for the
generated task parameters and calculate the second error, using the
same HL-model. Here we assume that it ``practiced'' with the selected
mode (see lines 11-15 of the Algorithm~\ref{alg:ts_error_valc}). We
then calculate the utility of this practice by comparing the two
errors (pre- and post-practice) and add the new data point to our GP
model. We apply noise to the difficulty of the task parameters as well
as the final utility to account for the variability of learners in the
respective group, differences within the set of randomly generated
tasks within the complexity level, as well as the variance in
performance of the learner depending on factors such as focus,
fatigue, etc.

\section{Results}

\begin{figure}
\centerline{\includegraphics[width=0.99\columnwidth]{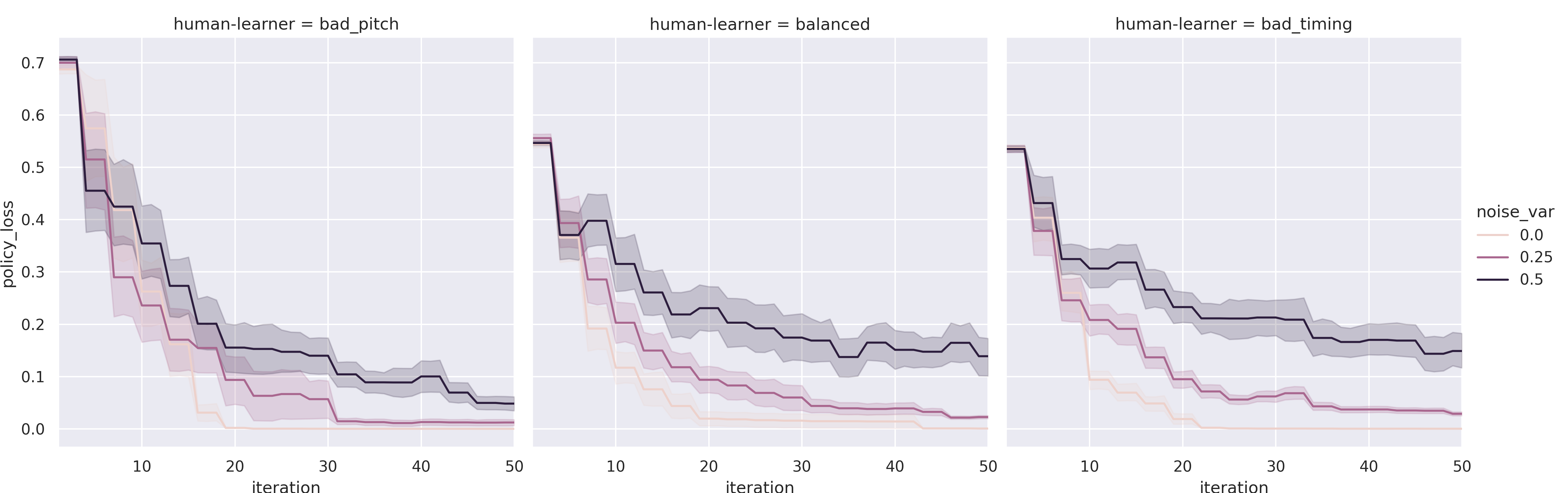}	}
\caption{The convergence towards the optimal policy for each of the
  \textit{HLs} with their respective weaknesses. Each combination was
  run 27 times, average and standard deviation are shown.}
\label{fig:ts_performers}
\end{figure}

We train the GP by iterating the above algorithm (see
Algorithm~\ref{alg:ts_experiment}). For implementation we have used
gpyopt library~\cite{gpyopt}.  Figure~\ref{fig:ts_performers}
illustrates convergence of the training process for three different
groups of HLs. Each plot illustrates results for different types of
noise added to the calculated utility function.  We evaluate the
learning progress of our model (depicted on the $y$-axis) based on the
policy-loss (see Equation~\ref{eqn:policy-loss}) calculated based on
our prior knowledge of the ground truth policy.
We  calculate loss that,
for a given task-parameter tuple, describes the missed utility of
selecting a certain practice mode (zero if the optimal mode is
selected) as follows: $\text{loss} = u^\text{optimal} - u^\textit{selected}$ and
$\text{loss}_\text{max} = u^\text{optimal} - u^\textit{non-optimal}$.
 This can be applied to the whole policy by summing over all
considered parameter combinations  and dividing by the total possible loss.
For 
$T = \{(\text{bpm}, \text{note\_range}) | \text{bpm} \in \{50, \ldots,
200\}, \text{note\_range} \in \{0,1,2\}\}$  we define the policy-loss as follows:
\begin{equation}
   \label{eqn:policy-loss} 
  \text{policy-loss} = \frac{\sum_T \text{loss}} {  \underset{T}{\text{median }} \text{loss}_\text{max} \times |T| }.
  \end{equation}

\begin{figure}
\begin{subfigure}{.5\textwidth}
  \centering
  \includegraphics[width=0.9\linewidth]{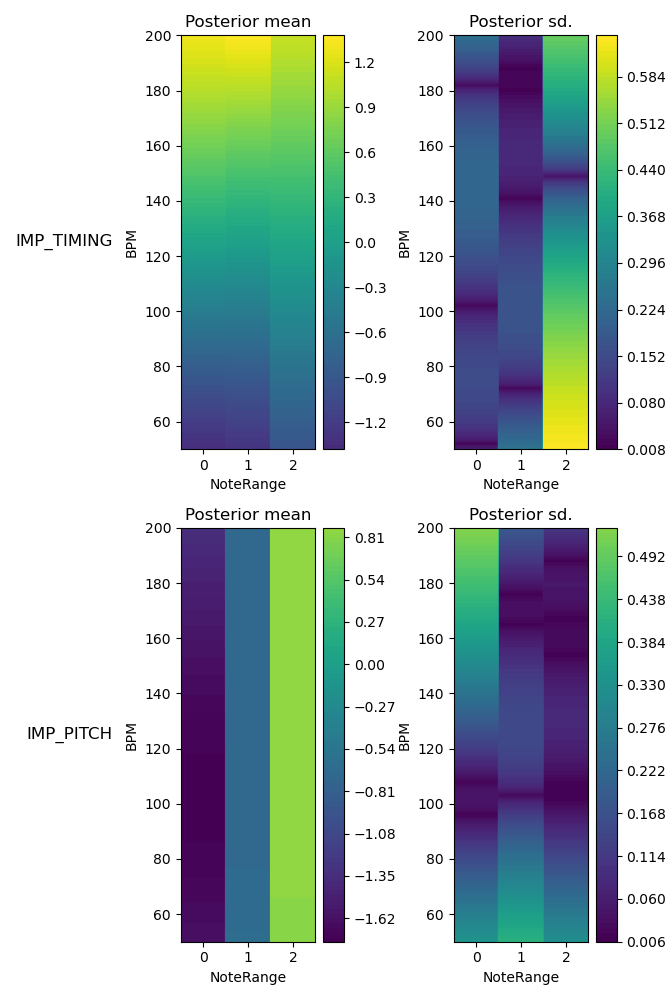}  
  \caption{No noise - optimal policy}
  \label{fig:ts_detailed_optimal}
\end{subfigure}
\begin{subfigure}{.5\textwidth}
  \centering
  \includegraphics[width=0.9\linewidth]{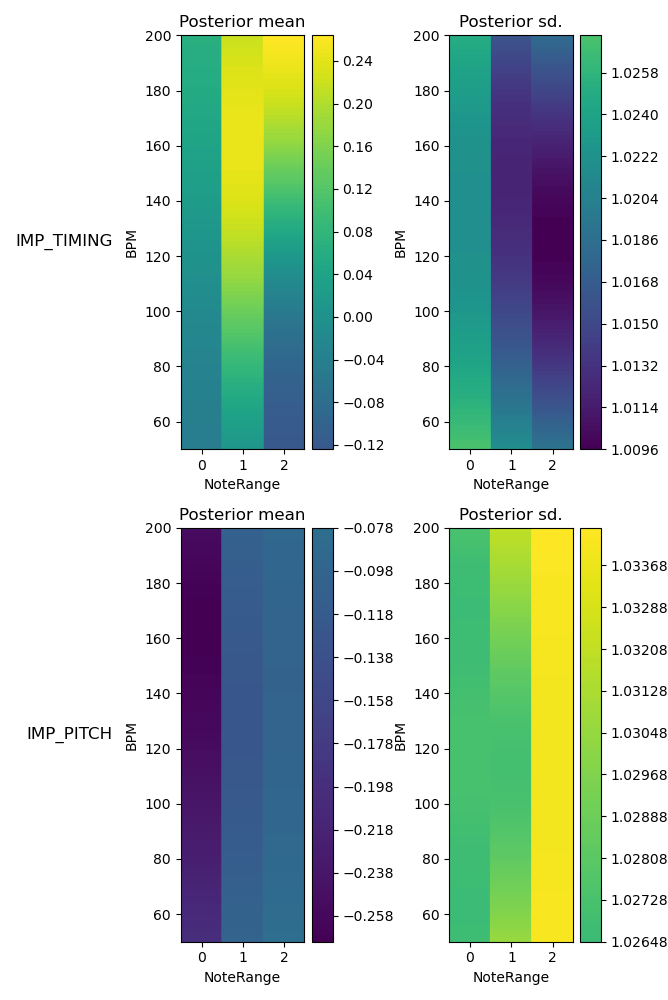}  
  \caption{0.5 noise std - 16\% policy-loss}
  \label{fig:ts_detailed_noise}
\end{subfigure}	

\caption{Visualization of the internal states of two GPs after 20 data-points each. Mean and standard deviation for both practice modes are shown for each process. The left GP had no noise applied to its data and learned the optimal policy (for the balanced learner) and dependencies (timing utility based only on bpm; pitch utility on note-range). The right GP had noise with a standard deviation of 0.5 applied during the creation of its input data. It fails to cleanly make out the true utilities and dependencies, and its resulting policy-loss is 16\%.
Each policy is the \textit{argmax} of the two practice-mode means.
}
\label{fig:ts_detailed_total}
\end{figure}

Adding noise into the system, equally applied to the initial
difficulty of the task, as well as the utility of a practice mode,
leads to slower convergence towards the optimal policy, as can be seen
in Figure~\ref{fig:ts_performers}. It is nevertheless clear that
learning takes place and no overfitting artifacts appear.  Figure
\ref{fig:ts_detailed_total} gives more insight into how the noise
hinders the regression towards the \textit{true} utilities.
Figure~\ref{fig:ts_detailed_total} visualizes the trained GPs with and without noise added to the utility of the practice mode.

\section{Discussion}

In this work we have introduced the concept of \textit{practice modes} and
presented our approach to scaffolding the learner by dynamic practice
mode selection, designed to optimize learner's progress in learning
how to play piano.  On an example including a large set of simulated
tasks on a single complexity level, two simulated practice modes and three
groups of human learners, we have illustrated the functionality of a
utility-based scaffolding infrastructure. One future work direction
will be to perform a study with real human learners with the goal to
verify the efficacy of our approach. Another will consist of
integrating Hidden Markov Model (HMM)-based methods as described
in~\cite{Mu2020TSA} to determine when or whether to move the human
learner up and down the curriculum tree, i.e. from one complexity
level to another to better model longer-term learning
processes. 

We believe that the proposed approach to scaffolding, illustrated on
the specific domain of piano practice, may be extended to other motor
tasks, including sports and rehabilitation, where complex skills also
need to be learned.  There is mixed evidence in the field of motor
learning as to whether learning a complex skill is better performed on
the task as a whole, or whether and how it is better to divide the
task into parts and learn these parts individually before combining
them \cite{schmidt05,fontana09}. The question
that is asked in these studies is whether the learning of the parts
performed individually transfers effectively to the situation when the
parts are combined.  A very general approach to both task 
and practice mode definition, which enables us to define a practice
mode by any combination of the ``parent task'' attributes, will give us
a strong formal framework to address the open research
questions of whether and how to optimally learn (``part'' vs. ``whole''
practice) for each individual learner.
The combination of an individually optimized
training program on one complexity level, combined with appropriately
moving between complexity levels (as described above)
may provide beneficial features from both part and whole learning
including the optimal time for switching complexity levels.

The use of automated techniques for individualizing motor learning as used in this study has
significant potential \cite{santos16} for improving learner
outcomes. As different learners show different learning curves, the
individual customization of motor training can ensure that each
learner receives an appropriate next exercise to maximize the amount
of learning taking place in a given amount of time. Additionally, maintaining appropriate difficulty
levels seems to be an important factor in determining intrinsic motivation for a task \cite{lomas17}.
Further, our learning infrastructure can be extended for measuring the
effect of factors such as augmented feedback \cite{mozer2019} and reward
\cite{codol2020} on motor learning at an individual level, and for
determining the optimal presentation of feedback and reward for
improving motor learning.
The interaction of feedback, reward and individually optimized learning tasks may lead to further enhancement
of the motor learning process.

\subsection{Limitations}

Currently our GP maps to a single output dimension: utility; defined
as the sum of improvements across all error aspects. This procedure is
reasonable for our simulated experiments, where the hyperparameters
were chosen to lead to three different policies in the end (see Figure
\ref{fig:ts_optimal_policies}). In the real world, the error
differences come in different modalities (e.g. seconds, half-notes off
pitch) and it will not be sufficient to simply add them together. It is
difficult to compare different expected improvement aspects and weigh
them against one another. 
One solution would be to bin the improvements per aspect, for example a one
second timing improvement over 7 bars, into categories like \textit{no
  improvement}, \textit{slight improvement}, \textit{improvement},
\textit{large improvement}. These bins could be defined by
semi-experts, describing the magnitude of improvement between two real
world data-samples.  Determining an intuitive weighting algorithm
based on such bins should be more feasible.

The current approach does not address the question of the number of
repetitions necessary to master the task specified by the practice
mode. Finding an optimal number of repetitions is an interesting
extension of this work which can be achieved in a similar way to~\cite{Mu2020TSA}
by tracking ``not mastered/engaged'', ``not mastered/not engaged'' and
``mastered'' states during practice.  We define an optimal practice
schedule as one that maximizes engagement, which  is likely to results in a
higher persistence of the human learner.

In the beginning of human learner experiments, very few data points
exist, and therefore a robust utility estimation for a new human
learner is not possible without their own GP training. This needs to
be approached by gathering more data from different human learners,
and allocating them to a growing set of groups according to their
attributes. Once sufficient data has been acquired, a new learner
could be allocated to their own group by a dynamic adjustment
procedure, evaluating their individual attributes.  Furthermore, we
will explore the idea of forming a meta-model based on multiple human
learners, which then during practice adjusts to the individual
learner's needs.  This feature is in general necessary due to the fact
that the learner may change their attributes as they practice over
long periods of time. Over short periods of time, the learning
progress of a HL may be accounted for by an extension of the GP 
 to learning  non-stationary system dynamics, such as described
in~\cite{Rottmann2010LNS}. 

Another possible improvement may be made to the music score
generation. Although it is advantageous to be completely automated for
an input of task attributes specific to a given complexity level, it
produces random note combinations which may not be very motivating for
the HLs. For the experiments with real HLs, we will use wave function
collapse \cite{kim19} to generate pieces that are different to known
tunes, but through a slight similarity with the original may be more
pleasant to the learners. Another option that will enable us to create
music scores with a similarity to the known original tune is to
additionally anchor them in the harmonic progression of the original.
An adherence to  basic musical
rules would guarantee a sufficiently musical outcome. e.g. formulaic
cadences, and overall note distributions that reflect established
tonal hierarchies~\cite{Krumhansl2010TTH}.

\begin{ack}
  This work is supported by a Grant from the GIF, the German-Israeli
  Foundation for Scientific Research and Development. Luisa Zintgraf
  is supported by the 2017 Microsoft Research PhD Scholarship Program,
  and the 2020 Microsoft Research EMEA PhD Award. We are thankful to
  the music experts Hila Tamir Ostrover, Yuval Carmi, Sigal Meiri, and
  Ronit Seter for their inspiring contribution that provided a
  foundation for this work.
  \end{ack}
\bibliographystyle{plainnat}
\bibliography{references}

\appendix
\section{Appendix}
\subsection{Music Score Generation  for Tasks and Practice Modes}
\label{appendix:task_generation}
\begin{figure}[h!]
\centerline{\includegraphics[width=0.95\columnwidth]{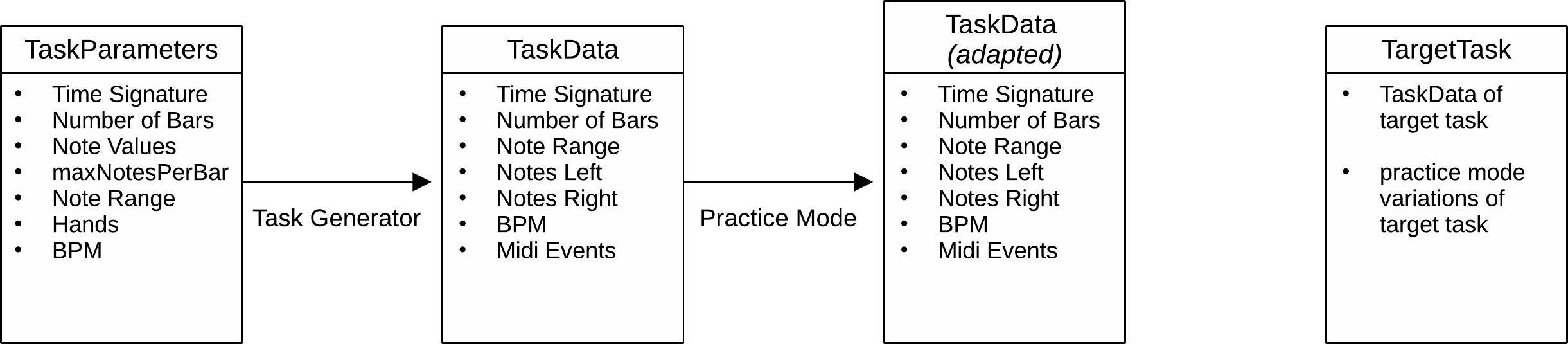}	}
\caption{Overview of the underlying data structures. Every tasks
  starts with \texttt{TaskParameters} which then get passed into a
  generator to render a \texttt{TaskData} object which includes the
  score (represented by the notes in \texttt{Notes Left, Notes
    Right}). This process is probabilistic and thus a large number of
  scores/\texttt{TaskData} objects can be generated from the same
  \texttt{TaskParameters}. Applying a \texttt{PraciceMode} to a
  \texttt{TaskData} object produces a new \texttt{TaskData} object
  with e.g. adapted notes. A \texttt{TargetTask} is an overarching
  construct that includes the \texttt{TaskData} of the main task, as
  well as the practice mode variations of said data.  }

\label{fig:ts_datastruct}
\end{figure}

Music scores are generated  by e.g. sampling from ranges defined by task parameters like note range. 
Music scores generated for practice modes in some cases need to be adapted to the practice mode, e.g in order to improve timing, every note in the score is changed to one single value such as middle C to allow the learner to concentrate on timing only, and not mind the pitch (see Figure~\ref{fig:ts_datastruct}). 

\subsection{GUI}
\label{appendix:piano_gui}
\begin{figure}[h!]
  \centering
  \includegraphics[width=\linewidth]{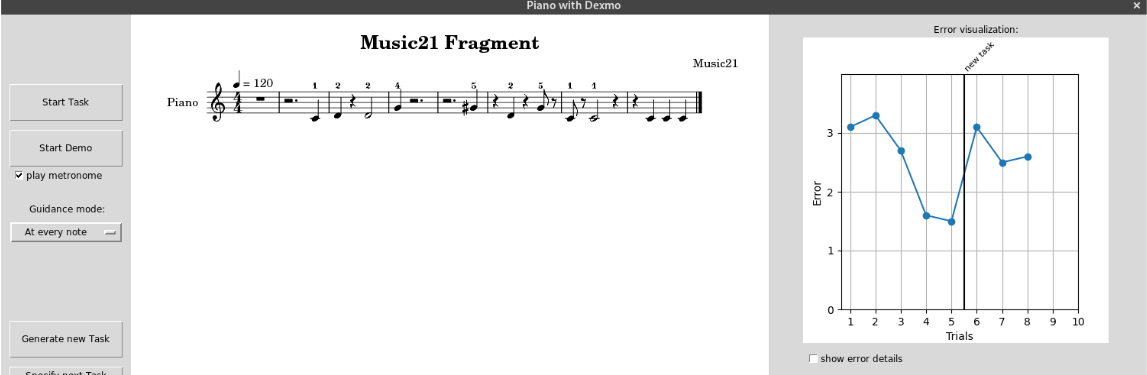}
  \caption{\label{fig:gui} Left: an example of a score piece randomly
    generated by the music score generator based on the specified task
    parameters, which is presented to a study participant via
    GUI. Both the notes as well as the fingering can be generated
    automatically (see \cite{pianoplayer}). Right: a simple
    error visualization.  }
\end{figure}

Figure \ref{fig:gui} shows an example of a task presented to the HL in real setting (which is part of future work). 

\end{document}